\documentclass[reprint,amsmath,amssymb,aps,prfluid]{revtex4-1}  

\usepackage{graphicx}
\usepackage{dcolumn}
\usepackage{bm}
\usepackage{multirow} 

\def\be{\begin{equation}} 
\def\ee{\end{equation}} 
\def\intd{\,\mathrm{d}} 

\begin{document}

\title{Cascades and Spectra of a Turbulent Spinodal Decomposition in 2D Symmetric Binary Liquid Mixture}
\author{Xiang Fan}\affiliation{University of California at San Diego, La Jolla, California 92093}
\author{P. H. Diamond}\affiliation{University of California at San Diego, La Jolla, California 92093}
\author{L. Chac\'on}\affiliation{Los Alamos National Laboratory, Los Alamos, New Mexico 87545}
\author{Hui Li}\affiliation{Los Alamos National Laboratory, Los Alamos, New Mexico 87545}
\date{\today} 

\begin{abstract}
We study the fundamental physics of cascades and spectra in 2D Cahn-Hilliard-Navier-Stokes (CHNS) turbulence, and compare and contrast this system with 2D MagnetoHydroDynamic (MHD) turbulence. The important similarities include basic equations, ideal quadratic invariants, cascades and the role of linear elastic waves. Surface tension induces elasticity, and the balance between surface tension energy and turbulent kinetic energy determines a length scale (Hinze scale) of the system. The Hinze scale may be thought of as the scale of emergent critical balance between fluid straining and elastic restoring forces. The scales between the Hinze scale and dissipation scale constitute the elastic range of the 2D CHNS system. By direct numerical simulation, we find that in the elastic range, the mean square concentration spectrum $H^\psi_k$ of the 2D CHNS system exhibits the same power law ($-7/3$) as the mean square magnetic potential spectrum $H^A_k$ in the inverse cascade regime of 2D MHD. This power law is consistent with an inverse cascade of $H^\psi$, which is observed. The kinetic energy spectrum of the 2D CHNS system is $E^K_k\sim k^{-3}$ if forced at large scale, suggestive of the direct enstrophy cascade power law of 2D Navier-Stokes (NS) turbulence. The difference from the energy spectra of 2D MHD turbulence implies that the back reaction of the concentration field to fluid motion is limited. We suggest this is because the surface tension back reaction is significant only in the interfacial regions. The interfacial regions fill only a small portion of the 2D CHNS system, and their interface packing fraction is much smaller than that for 2D MHD.
\end{abstract}
\maketitle

\section{Introduction}

Binary liquid mixtures can pass spontaneously from one miscible phase to two coexisting immiscible phases following a temperature drop. This second-order phase transition is called a spinodal decomposition. The Cahn-Hilliard-Navier-Stokes (CHNS) model \cite{cahn_free_1958,cahn_spinodal_1961} is the standard model for binary liquid mixture undergoing spinodal decomposition. The 2D CHNS system is as follows: (the definitions and derivation are discussed below)
\begin{align}
\partial_t\psi+\mathbf{v}\cdot\nabla\psi&=D\nabla^2(-\psi+\psi^3-\xi^2\nabla^2\psi) \label{CHNS1}\\
\partial_t\omega+\mathbf{v}\cdot\nabla\omega&=\frac{\xi^2}{\rho}\mathbf{B}_\psi\cdot\nabla\nabla^2\psi+\nu\nabla^2\omega \label{CHNS2}\\
\mathbf{v}=\mathbf{\hat{z}}\times\nabla\phi&,\ \omega = \nabla^2\phi \label{CHNS3}\\
\mathbf{B}_\psi=\mathbf{\hat{z}}\times\nabla\psi&,\ j_\psi = \xi^2\nabla^2\psi \label{CHNS4}
\end{align}
The definitions of the variables are discussed later in the paper. It is evident that this system is closely analogous to the 2D MagnetoHydroDynamics (MHD) model for plasmas:
\begin{align}
\partial_t A+\mathbf{v}\cdot\nabla A&=\eta\nabla^2 A \label{MHD1}\\
\partial_t\omega+\mathbf{v}\cdot\nabla\omega&=\frac{1}{\mu_0\rho}\mathbf{B}\cdot\nabla\nabla^2 A+\nu\nabla^2\omega \label{MHD2}\\
\mathbf{v}=\mathbf{\hat{z}}\times\nabla\phi&,\ \omega = \nabla^2\phi \label{MHD3}\\
\mathbf{B}=\mathbf{\hat{z}}\times\nabla A&,\ j = \frac{1}{\mu_0}\nabla^2 A \label{MHD4}
\end{align}
Since 2D MHD turbulence has been well studied \cite{diamond_self-consistent_2005,diamond_modern_2010,pouquet_two-dimensional_1978,pouquet_strong_1976,celani_active_2002,celani_active_2004,biskamp_magnetohydrodynamic_2003,biskamp_two-dimensional_2001,biskamp_dynamics_1994,kraichnan_inertial-range_1965,iroshnikov_turbulence_1964,servidio_magnetic_2009,matthaeus_rapid_2008}, it provides us with potential insight and guidance for exploring the physics of 2D CHNS turbulence. The comparison of 2D MHD and the 2D CHNS system is shown in Table~\ref{comparison}, the details are discussed later in this paper.

\begin{table*}
\caption{Comparison of 2D MHD and the 2D CHNS system.}
\scriptsize
\begin{center}
\begin{tabular}{ccc}
\hline
\hline
& 2D MHD & 2D CHNS\\
\hline
Ideal Quadratic Conserved Quantities & Conservation of $E$, $H^A$ and $H^C$ & Conservation of $E$, $H^\psi$ and $H^C$\\
Role of elastic waves & Alfven wave couples $\mathbf{v}$ with $\mathbf{B}$ & CHNS linear elastic wave couples $\mathbf{v}$ with $\mathbf{B}_\psi$\\
Origin of elasticity & Magnetic field induces elasticity & Surface tension induces elasticity\\
Origin of the inverse cascades & The coalescence of magnetic flux blobs & The coalescence of blobs of the same species\\
The inverse cascades & Inverse cascade of $H^A$ & Inverse cascade of $H^\psi$\\
Power law of spectra & $H^A_k\sim k^{-7/3}$ & $H^\psi_k\sim k^{-7/3}$\\
\hline
\hline
\end{tabular}
\end{center}
\label{comparison}
\end{table*}

The similarity between binary liquid mixture and 2D MHD was first discussed by Ruiz and Nelson \cite{ruiz_turbulence_1981}. They addressed only the regime when the binary liquid mixture is miscible, i.e. \textit{above} the critical temperature. The governing equation for this regime is
\be
\partial_t\psi+\mathbf{v}\cdot\nabla\psi=D\nabla^2 \psi
\ee
In this limit, basically there is no difference from 2D MHD. However, the more interesting and challenging regime occurs when the binary liquid mixture undergoes spinodal decomposition, i.e. \textit{below} the critical temperature.

\begin{figure}[htbp]
   \centering
   \includegraphics[width=\columnwidth]{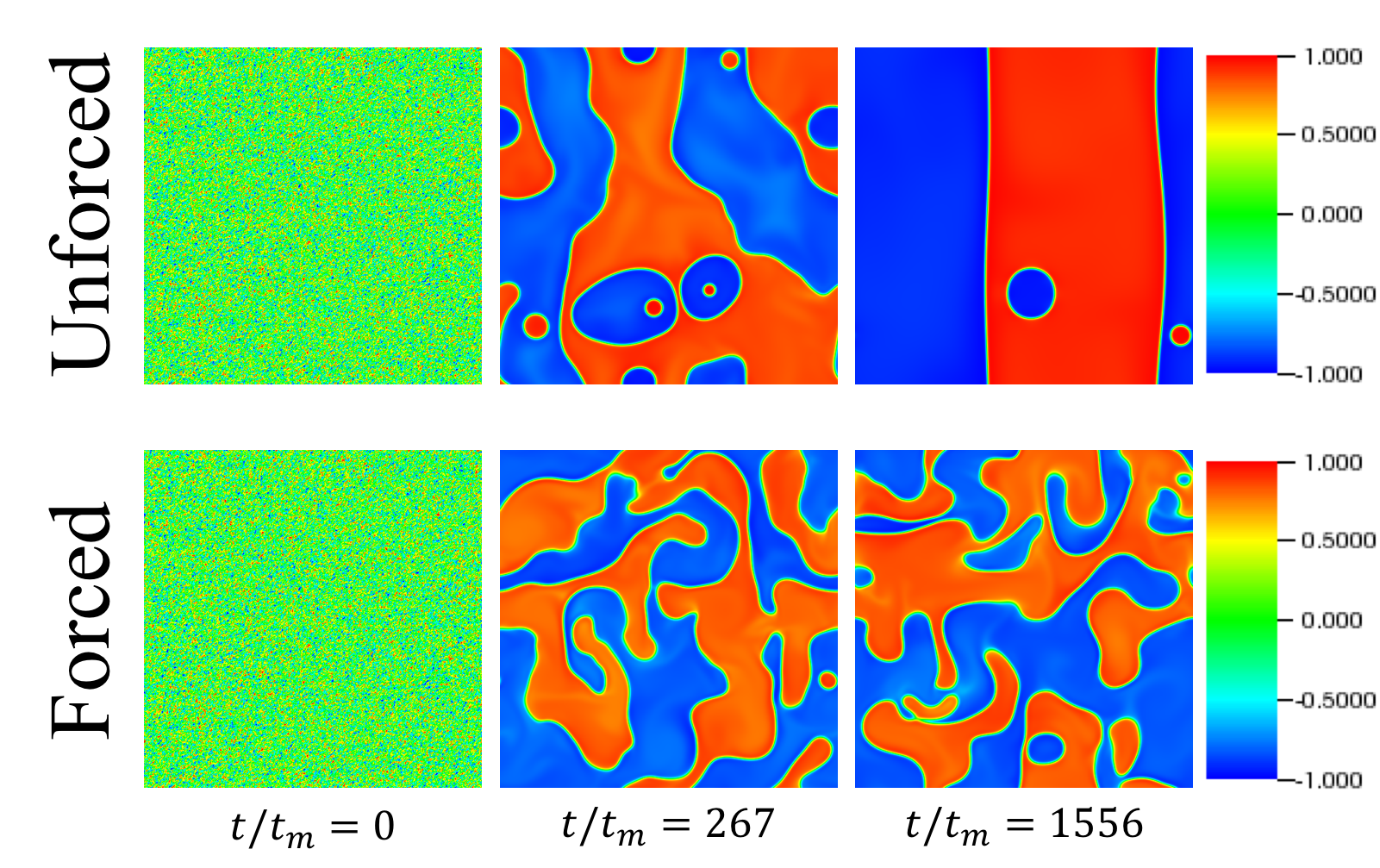} 
   \caption{Top panels are pseudo color plots of $\psi$ field for an unforced run (Run1) at various times; bottom panels are the ones for a forced run (Run4). Time $t$ is normalized by the diffusive mixing time $t_m=\xi^2/D$.}
   \label{pcolor}
\end{figure}

When the binary liquid mixture is quenched below the critical temperature, spinodal decomposition occurs. Small scale blobs tend to coalesce and form larger blobs \cite{kendon_inertial_2001,kendon_3d_1999,berthier_phase_2001,siggia_late_1979}, see Fig.~\ref{pcolor} (top panel) for an illustration. The blob size grows as $L\sim t^{2/3}$ if unforced \cite{furukawa_spinodal_2000}. The length scale growth can be arrested by external fluid forcing, and an emergent characteristic length scale of the blob size is formed by the critical balance between turbulent kinetic energy and surface tension energy in 2D CHNS turbulence \cite{berti_turbulence_2005}. In 3D, the length scale growth is also arrested when proper external forcing is applied, and the emergent characteristic length scale of blob size is consistent with the Hinze scale: $L_H\sim(\frac{\rho}{\sigma})^{-3/5}\epsilon^{-2/5}$ where $\rho$ is density, $\sigma$ is surface tension, and $\epsilon$ is the energy dissipation rate per unit mass \cite{perlekar_spinodal_2014,perlekar_droplet_2012}. In the inverse energy cascade regime of the 2D CHNS system, the characteristic length scale is also consistent with the Hinze scale \cite{perlekar_two-dimensional_2015}. 

Previous studies did not adequately separate the Hinze scale from the dissipation scale. We define the elastic range as the range of scales from the Hinze scale down to the dissipation scale. This is where the surface tension induced elasticity is important to the dynamics. The 2D CHNS system is more MHD-like in the elastic range. The power laws of the turbulent spectra in the elastic range were not investigated by previous studies.

In this study, we first describe the fundamental theory of spinodal decomposition in Sec.~\ref{sec2}. In Sec.~\ref{sec3}, we compare and contrast 2D CHNS with 2D MHD in terms of basic equations, ideal quadratic conserved quantities, cascades, and linear elastic wave. The concepts of the Hinze scale and the elastic range are explained in detail in Sec.~\ref{sec4}. Next we use the PIXIE2D code \cite{chacon_implicit_2002,chacon_2d_2003} to simulate the 2D CHNS system in Sec.~\ref{sec5}. We focus on the turbulent spectra and cascades in the elastic range, and compare them with 2D MHD. Conclusions and discussions are presented in Sec.~\ref{sec6}.

\section{Governing Equations for Spinodal Decomposition}\label{sec2}
We consider spinodal decomposition in a symmetric (50\%-50\%) binary liquid mixture of equal density. Spinodal decomposition is a second-order phase transition, and so can be modeled by Landau theory. 

\begin{figure}[htbp]
   \centering
   \includegraphics[width=0.6\columnwidth]{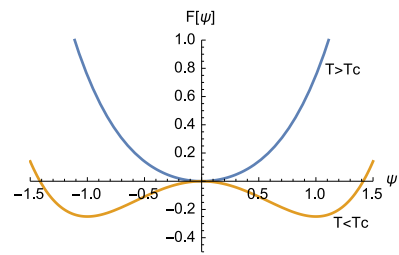} 
   \caption{Free energy functional $F[\psi]$ for $T>T_c$ and $T<T_c$.}
   \label{fpsi}
\end{figure}

The corresponding order parameter is the local relative concentration $\psi(\mathbf{x},t)$:
\be
\psi = \frac{\rho_A-\rho_B}{\rho_A+\rho_B}
\ee
where $\rho_A$ and $\rho_B$ are the local densities of the two species. When $\rho_B=0$, $\psi=+1$ implies an A-rich phase; when $\rho_A=0$, $\psi=-1$ implies a B-rich phase. The range of $\psi$ is thus $\psi\in[-1,1]$. The free energy functional reads as:
\be
F[\psi]=\int (\frac{1}{2}A\psi^2+\frac{1}{4}B\psi^4+\frac{\xi^2}{2}|\nabla\psi|^2)\intd\mathbf{r}
\ee
where $A$ and $B$ are coefficients of a Taylor expansion, and $\xi$ is a coefficient describing the strength of the surface tension interaction. $\xi$ also characterizes the interfacial thickness. The first two terms characterize the second-order phase transition dynamics, while the last term is the curvature penalty. In Landau theory, $B$ must always be greater than $0$ for the system to be thermodynamically stable, while $A$ can be either positive or negative, i.e.:
\be
A=A_0(T-T_c)
\ee
where $A_0$ is some temperature independent constant, $T$ is the temperature and $T_c$ is the critical temperature for spinodal decomposition. As shown in Fig.~\ref{fpsi}, when $T>T_c$, $A>0$, the free energy $F[\psi]$ has a ``V" shape, so there is only one minimum at $\psi=0$. When $T<T_c$, $A<0$, the free energy $F[\psi]$ has a ``W" shape, so there is one unstable maximum at $\psi=0$, and two minima at $\psi=\pm\sqrt{-\frac{A}{B}}$. When the homogeneous phase of the binary liquid mixture is quenched down to below the critical temperature, the $\psi=0$ phase becomes unstable because the system tends to reach its minimal energy, and the system now prefers the $\psi=\pm\sqrt{-\frac{A}{B}}$ phases, implying phase separation. Because of the definition of $\psi$, the minimal energy should be reached when $\psi=\pm1$, so we have $B=-A$. For simplicity, we study the isothermal case when the temperature is fixed below $T_c$, i.e. $A$ is constant. Without loss of generality, we set $B=-A=1$:
\be
F[\psi]=\int (-\frac{1}{2}\psi^2+\frac{1}{4}\psi^4+\frac{\xi^2}{2}|\nabla\psi|^2)\intd\mathbf{r}
\ee

The dynamics of the binary liquid mixture under spinodal decomposition is fully determined by this free energy functional. The chemical potential is 
\be
\mu=\frac{\delta F}{\delta\psi}=-\psi+\psi^3-\xi^2\nabla^2\psi
\ee
According to Fick's Law $\mathbf{J}=-D\nabla\mu$ (where $D$ is diffusivity) and the continuity equation $\mathrm{d}\psi/\mathrm{d}t+\nabla\cdot\mathbf{J}=0$, we obtain the Cahn-Hilliard Equation:
\be
\mathrm{d}\psi/\mathrm{d}t=D\nabla^2(-\psi+\psi^3-\xi^2\nabla^2\psi)
\ee
The total derivative is $\mathrm{d}/\mathrm{d}t=\partial\psi/\partial t+\mathbf{v}\cdot\nabla$ when flow is present, where $\mathbf{v}$ is velocity. The fluid motion satisfies Navier-Stokes Equation, with an additional force term due to surface tension:
\be
\partial_t\mathbf{v}+\mathbf{v}\cdot\nabla\mathbf{v}=-\frac{1}{\rho}\nabla p-\frac{\xi^2}{\rho}\nabla^2\psi\nabla\psi+\nu\nabla^2\mathbf{v}\label{NS_with_fs}
\ee 
Here $\nu$ is viscosity, $p$ is pressure, and $\rho=\rho_A+\rho_B$ is density. The second term on the R.H.S. comes from the surface tension force, which has the from $-\frac{1}{\rho}\psi\nabla\mu$. This means that the force pushes two species in opposite directions, with a strength proportional to the gradient of the chemical potential. This surface tension force can be written in the form $-\frac{1}{\rho}\nabla(-\frac12\psi^2+\frac34\psi^4-\xi^2\psi\nabla^2\psi)-\frac{\xi^2}{\rho}\nabla^2\psi\nabla\psi$. The first part can be absorbed into the definition of pressure $p$, leaving the second part as in Eq.~(\ref{NS_with_fs}). Finally, for 2D incompressible flow, $\nabla\cdot\mathbf{v}=0$, so it is more convenient to take the curl of Eq.~(\ref{NS_with_fs}) and work with the vorticity equation. 

To summarize, the governing equations for spinodal decomposition in 2D symmetric binary liquid mixture are the Cahn-Hilliard-Navier-Stokes (CHNS) equations: Eqs.~(\ref{CHNS1}) - (\ref{CHNS4}), where $\phi$ is the stream function, $\omega$ is vorticity, and $\mathbf{B}_\psi$ and $j_\psi$ are analogous to magnetic field and current in MHD, respectively, which will be discussed in the next section.

\section{Comparison and contrast of 2D CHNS Turbulence and 2D MHD Turbulence}\label{sec3}

\subsection{Basic Equations}

The 2D CHNS system is an analogue to 2D Magnetohydrodynamics (MHD) in plasma physics. MHD turbulence is comparatively better understood due to several decades of extensive study. By comparison and contrast of 2D CHNS turbulence and 2D MHD turbulence, we can understand each more clearly. 

The 2D MHD equations are Eqs.~(\ref{MHD1}) - (\ref{MHD4}), where $A$ is the scalar magnetic potential in 2D, $\mathbf{B}$ is magnetic field, $j$ is current, $\eta$ is resistivity, and $\mu_0$ is magnetic permeability. Comparing Eqs.~(\ref{CHNS1}) - (\ref{CHNS4}) and Eqs.~(\ref{MHD1}) - (\ref{MHD4}), we immediately grasp the correspondence between these two systems, which is summerized in Table~\ref{correspondence}. Note that the surface tension force $\frac{\xi^2}{\rho}\mathbf{B}_\psi\cdot\nabla\nabla^2\psi$ in Eq.~(\ref{CHNS2}) and the $\mathbf{j}\times\mathbf{B}$ force $\frac{1}{\mu_0\rho}\mathbf{B}\cdot\nabla\nabla^2 A$ in Eq.~(\ref{MHD2}) have the same structure.

\begin{table}
\caption{The correspondence between 2D MHD and the 2D CHNS system.}
\begin{center}
\begin{tabular}{ccc}
\hline
\hline
& 2D MHD & 2D CHNS \\
\hline
Magnetic Potential & $A$ & $\psi$ \\
Magnetic Field & $\mathbf{B}$ & $\mathbf{B}_\psi$ \\
Current & $j$ & $j_\psi$ \\
Diffusivity & $\eta$ & $D$ \\
Interaction strength & $\frac{1}{\mu_0}$ & $\xi^2$ \\
\hline
\hline
\end{tabular}
\end{center}
\label{correspondence}
\end{table}%

The major difference is between the dissipation terms in Eq.~(\ref{CHNS1}) and Eq.~(\ref{MHD1}). The CHNS equations contain a negative diffusivity term $-D\nabla^2\psi$, self nonlinear diffusivity term $D\nabla^2\psi^3$ and a hyper-diffusivity term $-\xi^2 D\nabla^2\nabla^2\psi$. The MHD equations only contain one (positive) resistivity term $\eta\nabla^2 A$. Another difference to notice is that the concentration $\psi$ ranges from $-1$ to $1$, limited by physics definition $\psi = \frac{\rho_A-\rho_B}{\rho_A+\rho_B}$. The magnetic potential $A$ has no such restriction.

The CHNS system is more similar to MHD in 2D than in 3D, because magnetic potential $A$ is a scalar in 2D, but is a vector in 3D. The concentration $\psi$ is always a scalar, regardless of dimension.

\subsection{Ideal Quadratic Conserved Quantities}

The quadratic conserved quantities in the ideal system, which means $D,\eta=0$ and $\nu=0$ here, are important to the study of turbulent cascades. The real turbulent systems with finite dissipation are different from ideal systems, nevertheless, the ideal conserved quantities are still important constraints imposed on the nonlinear dynamics. In particular, the study of absolute equilibrium distributions of the ideal systems provides us indications of cascade directions.

It is known that there are 3 ideal quadratic conserved quantities in 2D MHD: total energy $E$ (which is the sum of kinetic energy $E^K$ and magnetic energy $E^B$), mean square magnetic potential $H^A$, and cross helicity $H^C$:
\begin{align}
E=E^K+E^B&=\int(\frac{\rho\mathbf{v}^2}{2}+\frac{\mathbf{B}^2}{2\mu_0})\intd^2x \\
H^A&=\int A^2\intd^2x \\
H^C&=\int\mathbf{v}\cdot\mathbf{B}\intd^2x
\end{align}
Note that $H^A$ is not a conserved quantity in 3D MHD; instead, the magnetic helicity $H^B=\int\mathbf{A}\cdot\mathbf{B}\intd^3x$ is conserved.

%
%
%

When the dissipation is set to $0$, the difference between the 2D CHNS system and 2D MHD disappears, so the ideal quadratic conserved quantities in the 2D CHNS system are the direct analogues of those in MHD, namely: total energy $E$, mean square concentration $H^\psi$, and cross helicity $H^C$:
\begin{align}
E=E^K+E^B&=\int(\frac{\rho\mathbf{v}^2}{2}+\frac{\xi^2\mathbf{B}_\psi^2}{2})\intd^2x \label{CHNS_energy}\\
H^\psi&=\int \psi^2\intd^2x \\
H^C&=\int\mathbf{v}\cdot\mathbf{B}_\psi\intd^2x
\end{align}

Note that some previous works \cite{berti_turbulence_2005,perlekar_two-dimensional_2015} use another definition of energy: $E'=E^K+F=\int(\frac{\rho\mathbf{v}^2}{2}-\frac{1}{2}\psi^2+\frac{1}{4}\psi^4+\frac{\xi^2}{2}|\nabla\psi|^2)\intd^2x$. This is also an ideal conserved quantity, but it is not \textit{quadratic}. In this paper, we focus on \textit{quadratic} conserved quantities, because higher-order conserved quantities are not strictly conserved when the $k$ space is discretized and truncated at large $k$. Since discretization and truncation are unavoidable when doing statistical physics and numerical simulation, only \textit{quadratic} conserved quantities are robust enough to be meaningful.

The physical meaning of cross helicity in the CHNS equations is not clear, as it is in MHD. The role of cross helicity is an interesting question, but it is beyond the scope of this paper. It will be investigated further in future works.

In addition, recall that there are only two ideal quadratic conserved quantities in 2D Navier-Stokes (NS) turbulence: kinetic energy $E^K$ and enstrophy $\Omega$:
\begin{align}
E^K&=\int\frac{\mathbf{v}^2}{2}\intd^2x \\
\Omega&=\int\frac{\omega^2}{2}\intd^2x
\end{align}
It is clear that the constraints on the dynamics of the CHNS system are more like those for 2D MHD than 2D NS. The conservation of enstrophy is broken in the 2D CHNS system by the surface tension force, just as it is broken by the $\mathbf{j}\times\mathbf{B}$ force in 2D MHD. Although enstrophy is not a strict ideal conserved quantity in 2D CHNS system, it is still useful to retain this concept, for reasons discussed below. 

\subsection{Cascades}

Turbulence cascade directions of various physics systems are suggested by the absolute equilibrium distributions, i.e. the Gibbs distribution \cite{biskamp_magnetohydrodynamic_2003,frisch_possibility_1975}. The peak of the absolute equilibrium distribution for each quadratic conserved quantity is a good indicator of the corresponding cascade direction. This approach only depends on the ideal quadratic conserved quantities of the system. Because the ideal quadratic conserved quantities of 2D CHNS and 2D MHD are direct analogues, we can then obtain an indication of the cascade directions in 2D CHNS by changing the name in variables. The summary of cascade directions of relevant physics systems are shown in Table~\ref{cascade_directions}.

The Gibbs distribution for 2D MHD is 
\be
\rho_G=Z^{-1}\exp(-\alpha E-\beta H^A-\gamma H^C)
\ee
where $\alpha$, $\beta$ and $\gamma$ are Lagrangian multipliers and $Z$ is the partition function. Similarly, the Gibbs distribution for 2D CHNS is
\be
\rho_G=Z^{-1}\exp(-\alpha E-\beta H^\psi-\gamma H^C)\label{rhoG}
\ee
By calculating each ideal spectral density from the above absolute equilibrium distribution, suggested cascade directions can be extracted. The second-order moment for a Gaussian distribution $\rho=Z^{-1}\exp{-\frac12\sum_{i,j}A_{ij}x_ix_j}$ is:
\be
\langle x_ix_j\rangle=A_{ij}^{-1}\label{Aij-1}
\ee
Write the ideal quadratic conserved quantities in terms of Fourier modes and in the discrete form, and restrict the index of summation $\mathbf{k}$ within the band $k_{min}<k<k_{max}$:
\begin{align}
E=\frac12&\sum_{\mathbf{k}}k^2(|\phi_{\mathbf{k}}|^2+|\psi_{\mathbf{k}}|^2)\\
H^\psi&=\sum_{\mathbf{k}}|\psi_{\mathbf{k}}|^2 \\
H^C&=\sum_{\mathbf{k}}k^2\phi_{\mathbf{k}}\psi_{\mathbf{-k}}
\end{align}
Plugging the above expressions into Eq.~(\ref{rhoG}) and Eq.~(\ref{Aij-1}) (set $\rho=1$ and $\xi^2=1$ for simplicity), it is then straightforward to obtain the expressions for ideal spectral densities:
\begin{align}
E^K_k&=\frac12 k^2\langle|\phi_{\mathbf{k}}^2|\rangle=\frac{2\pi k}{\alpha}(1+\frac{k^2\tan^2\theta}{k^2+(\beta/\alpha)\sec^2\theta})\\
E^B_k&=\frac12 k^2\langle|\psi_{\mathbf{k}}^2|\rangle=\frac{2\pi k}{\alpha}\frac{k^2\sec^2\theta}{k^2+(\beta/\alpha)\sec^2\theta}\\
H^\psi_k&=\langle|\psi_{\mathbf{k}}^2|\rangle=2k^{-2}E^B_k \\
H^C_k&=k^2\langle\phi_{\mathbf{k}}\psi_{-\mathbf{k}}\rangle=-\frac{2\gamma}{\alpha}E^B_k
\end{align}
where $\sin\theta=\gamma/(2\alpha)$. The requirement that $E^K_k$, $E^B_k$ and $H^\psi_k$ are always positive definite implies that $\alpha>0$, $k^2_{min}+(\beta/\alpha)\sec^2\theta>0$, and $|\gamma|<2\alpha$. If the spectrum is peaked at high $k$, and excitation is injected at intermediate scales, we expect the spectrum to relax towards high $k$ \cite{biskamp_magnetohydrodynamic_2003}. The trend suggests a direct cascade. Similarly, an inverse cascade is suggested if a spectrum is peaked at small $k$. So for the 2D CHNS system, we predict a \textit{direct} energy cascade and an \textit{inverse} cascade of $H^A_k$. The spectral transfer of cross helicity spectral density $H^C_k$ needs more consideration and is beyond the scope of this paper.

In 2D MHD, the inverse cascade of $H^A$ can be understood as the process of magnetic flux coalescence \cite{diamond_modern_2010}. Similarly, in 2D CHNS, the inverse cascade of $H^\psi$ can be related to the coalescence of blobs of the same species.

\begin{table}
\caption{The cascade directions for 2D MHD, CHNS and NS turbulences. }
\begin{center}
\begin{tabular}{ccc}
\hline
\hline
Physics System & Conserved Quantity & Cascade Direction\\
\hline
\multirow{2}{*}{2D MHD} & $E_k$ & Direct\\
& $H^A_k$ & Inverse\\
\hline
\multirow{2}{*}{2D CHNS} & $E_k$ & Direct\\
& $H^\psi_k$ & Inverse\\
\hline
\multirow{2}{*}{2D NS} & $E^K_k$ & Inverse\\
& $\Omega_k$ & Direct\\
\hline
\hline
\end{tabular}
\end{center}
\label{cascade_directions}
\end{table}%

\subsection{Linear Elastic Wave}

\begin{figure}[htbp]
   \centering
   \includegraphics[width=\columnwidth]{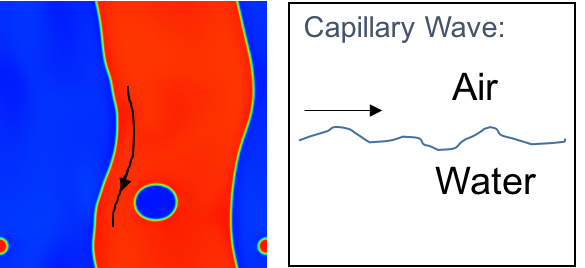} 
   \caption{The linear elastic wave (left) in the 2D CHNS system propagates only along the interface, similar to capillary wave (right).}
   \label{wave}
\end{figure}

Since Alfven waves play a crucial role in MHD turbulence, it is meaningful to examine the similar linear elastic wave in CHNS system. Recall that in the limit of small damping, the dispersion relation for the Alfven wave in 2D MHD is:
\be
\omega(\mathbf{k})=\pm\sqrt{\frac{1}{\mu_0\rho}}|\nabla A_0\times\mathbf{k}|-\frac12i(\eta+\nu)k^2
\ee
It is straightforward to linearize the CHNS equations and obtain a similar linear elastic wave:
\be
\omega(\mathbf{k})=\pm\sqrt{\frac{\xi^2}{\rho}}|\nabla\psi_0\times\mathbf{k}|-\frac12i(CD+\nu)k^2
\ee
where $C=[-1-6\psi_0\nabla^2\psi_0/k^2-6(\nabla\psi_0)^2/k^2-12\psi_0\nabla\psi_0\cdot i\mathbf{k}/k^2+3\psi_0^2+\xi^2k^2]$ is a dimensionless coefficient. The 2D CHNS system spontaneously leads to a state of phase separation. Inside a blob of the same species, the concentration field $\psi_0$ is homogeneous, so $\nabla\psi_0\rightarrow 0$. \textit{$\nabla\psi_0$ is large only along the interface of blobs}, as shown in Fig.~\ref{wave}. The CHNS linear elastic wave propagates along the interface of the two species where $\nabla\psi_0\neq0$, so it is much like a capillary wave. 

Alfven waves and CHNS linear elastic waves are similar, not only due to the resemblance of the dispersion relations, but also because both wave propagate along $\mathbf{B}_0$ or $\mathbf{B}_{\psi0}$ field lines. Both waves are elastic waves, in which magnetic tension and surface tension generate restoring forces that act as elasticity. The Alfvenization process in MHD turbulence couples $\mathbf{v}$ with $\mathbf{B}$, and even a weak mean magnetic field can spontaneously convert fluid eddies into Alfven waves \cite{diamond_modern_2010}. The Alfvenization process leads to Alfvenic equipartition $\rho\langle \mathbf{v}^2 \rangle \sim \frac{1}{\mu_0}\langle \mathbf{B}^2 \rangle$ of the fields. A similar elasticization process can also occur in the 2D CHNS system, because of the presence of linear elastic waves. The corresponding elastic equipartition for the 2D CHNS system is as follows:
\be
\rho\langle \mathbf{v}^2 \rangle \sim \xi^2\langle \mathbf{B}_\psi^2 \rangle\label{elastic_equipartition}
\ee

An interesting difference between Alfven wave and the CHNS linear elastic wave is that, the non-ideal part of the dispersion relation for CHNS linear elastic wave can be either positive or negative depending on $k$: if $CD+\nu>0$, then the wave is damped; but if $CD+\nu<0$, growth is possible. This wave growth is physical, and is responsible for the pattern formation during the linear phase, and the sustainment of sharp interfaces that separate phases during the dynamical evolution of the physical system in the nonlinear phase. It is important to note that treating this anti-diffusive term numerically is non-trivial, and requires unconditionally energy-stable temporal update schemes that ensure energy is either conserved or slightly dissipated. In this work, we have employed the MP-BDF2 energy-stable scheme proposed in Ref.~\cite{guillen-gonzalez_second_2014}, which in addition to being energy-stable, is unconditionally uniquely solvable.

\section{Important Length Scales and Ranges of 2D CHNS Turbulence}\label{sec4}

\begin{figure}[htbp]
   \centering
   \includegraphics[width=\columnwidth]{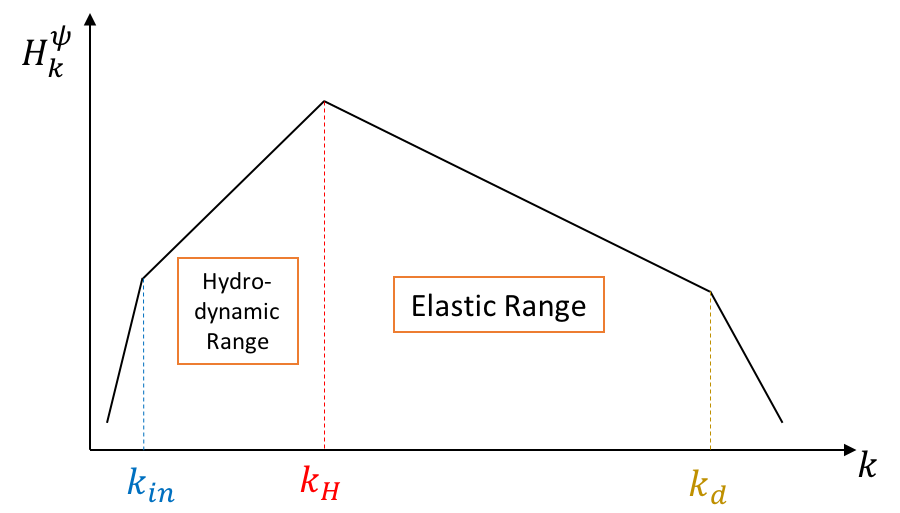} 
   \caption{The Hinze scale, hydrodynamic range and elastic range.}
   \label{ranges}
\end{figure}

\begin{figure}[htbp]
   \centering
   \includegraphics[width=0.5\columnwidth]{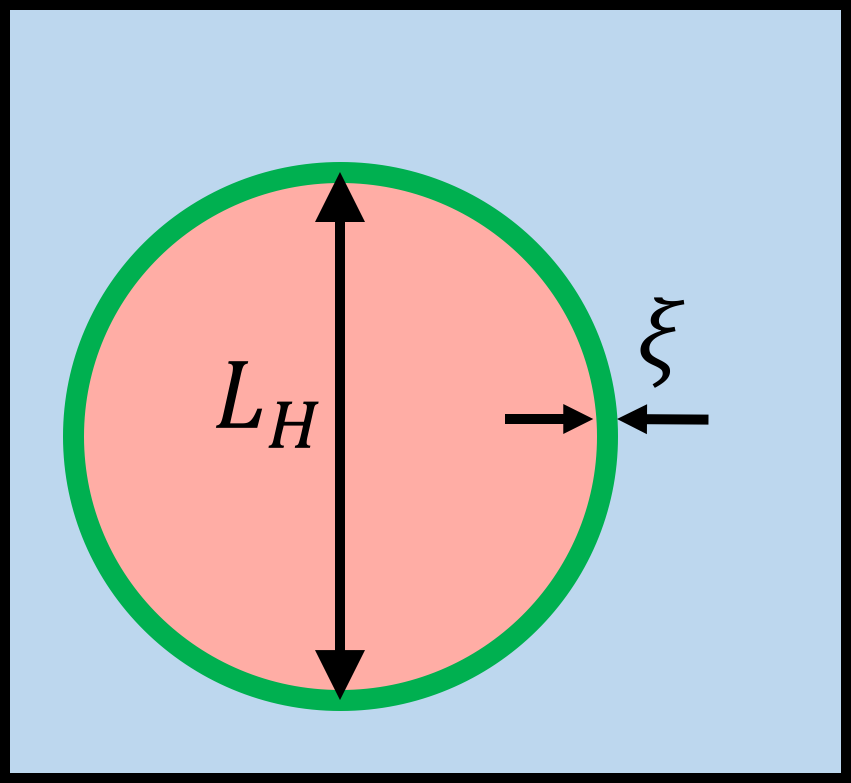} 
   \caption{The gradient length scales.}
   \label{blob_length_scales}
\end{figure}

In the forced 2D CHNS system, large blobs in the binary liquid mixture tend to be broken up by turbulent fluid straining, while small blobs tend to stick together due to surface tension. From this competition, a statistically stable length scale for the blob size, the Hinze scale $L_H$, emerges. $L_H$ is defined by balancing turbulent kinetic energy and surface tension energy \cite{hinze_fundamentals_1955, perlekar_spinodal_2014}:
\be
\frac{\rho\langle \mathbf{v}^2\rangle}{\sigma/L_H}\sim 1\label{Hinze_balance}
\ee
where $\sigma$ is surface tension. According to \cite{kendon_inertial_2001,kendon_3d_1999}, the surface tension is $\sigma=\sqrt{\frac89}\xi$. The surface tension energy can also be expressed in terms of $B_{\psi}^{rms}$ ($B_{\psi}^{rms}=\langle\mathbf{B}^2_\psi\rangle^{1/2}$). The key is to identify the relevant length scale for $\nabla\psi$. We propose to use the geometric mean of the blob size $L_H$ and the interface width $\xi$, because they are the longest and shortest gradient length scales, respectively, as shown in Fig.~\ref{blob_length_scales}. Assuming the length scale for $B_\psi^{rms}$ is the geometric mean of $L_H$ and $\xi$, i.e. $B_\psi^{rms}\sim\sqrt{\frac{\Delta\psi}{L_H}\frac{\Delta\psi}{\xi}}\sim\sqrt{\frac{1}{L_H\xi}}$, then the original expression for surface tension energy $\sigma/L_H$ is consistent with our expression $\xi^2\langle\mathbf{B}^2_\psi\rangle$ in Eq.~(\ref{CHNS_energy}). It is interesting to note that the critical balance Eq.~(\ref{Hinze_balance}) is then consistent with elastic equipartition ($\rho\langle \mathbf{v}^2 \rangle \sim \xi^2\langle \mathbf{B}_\psi^2 \rangle$).

The expression for the Hinze scale was originally derived for the 3D NS direct energy cascade regime \cite{hinze_fundamentals_1955}. The velocity was estimated using the Kolmogorov energy distribution law, $\langle \mathbf{v}^2\rangle/k_H\sim \epsilon^{2/3}k_H^{-5/3}$ where $\epsilon=\nu\int\omega^2\intd x^2$ is the kinetic energy dissipation rate per unit mass and $k_H=2\pi/L_H$. We then obtain the expression:
\be
L_H\sim(\frac{\rho}{\xi})^{-3/5}\epsilon^{-2/5}\label{Hinze_inv}
\ee
However, in the 2D NS direct enstrophy cascade regime, the velocity distribution is $\langle \mathbf{v}^2\rangle/k_H\sim \epsilon_\Omega^{2/3}k_H^{-3}$ where $\epsilon_\Omega=\nu\int(\nabla\times\omega\mathbf{\hat{z}})^2\intd x^2$ is the enstrophy dissipation rate per unit mass. Therefore, in 2D:
\be
L_H\sim(\frac{\rho}{\xi})^{-1/3}\epsilon_\Omega^{-2/9}\label{Hinze_dir}
\ee

Note that the Hinze scale depends on the magnitude of the external forcing via $\epsilon_\Omega$, and it does not depend on the scale of the external forcing. The Hinze scale separates the $k$ space into two ranges: the scales larger than $L_H$ form the hydrodynamic range, where the usual eddy break-up process dominates. The range of scales between $L_H$ and dissipation scale $L_d$ is the \textit{elastic range}, where the blob coalescence process dominates, as shown in Fig.~\ref{ranges}. Separation between the Hinze scale $L_H$ and dissipation scale $L_d$ is critical to defining an elastic range. The dissipation scale here should be related to the direct enstrophy cascade. By simple dimensional analysis, we obtain $L_d=(\nu^3/\epsilon_\Omega)^{1/6}$. Defining a dimensionless number for the ratio of $L_H$ to $L_d$ gives:
\be
L_H/L_d = Hd = (\frac{\rho}{\xi})^{-1/3}\nu^{-1/2}\epsilon_\Omega^{-1/18}
\ee
$Hd\gg 1$ is required to form a large enough elastic range. It is clear that reducing $\nu$ is an efficient way to obtain a longer elastic range.

\begin{figure}[htbp]
   \centering
   \includegraphics[width=\columnwidth]{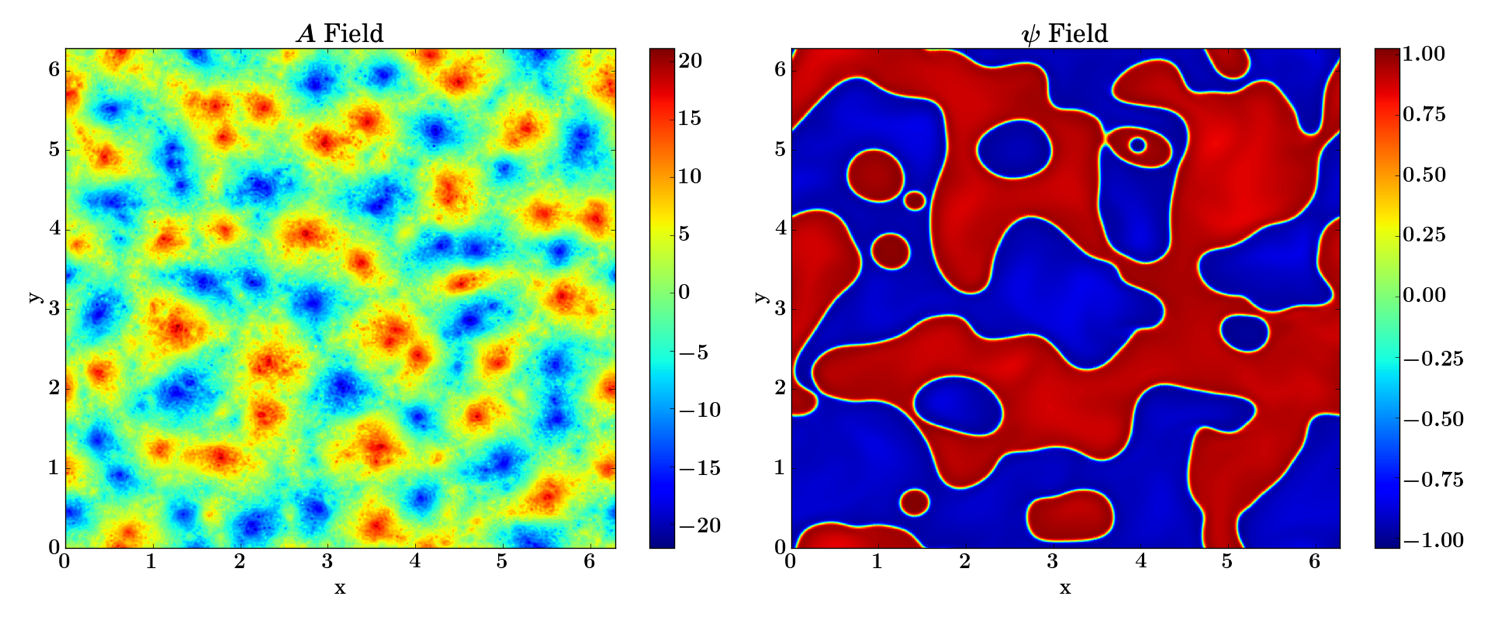} 
   \caption{The $A$ blobs in 2D MHD (Run6) and the $\psi$ blobs in the 2D CHNS system (Run2).}
   \label{apsi}
\end{figure}

The $A$ blobs in 2D MHD and $\psi$ blobs in the 2D CHNS system are shown side by side in Fig.~\ref{apsi}. In the elastic range of the 2D CHNS system, the blob coalescence process is analogous to the magnetic flux blob coalescence process in 2D MHD. The former leads to the inverse cascade of $H^\psi$, and the latter leads to the inverse cascade of $H^A$. In the elastic range of the 2D CHNS system, surface tension induces elasticity and plays a major role in defining a restoring force. Similarly, in 2D MHD, the magnetic field induces elasticity and make MHD different from a pure fluid. The 2D CHNS system is more MHD-like in the elastic range.

\section{Numerical Results}\label{sec5}

\subsection{Basic Setup}

We solve 2D CHNS Eqs.~(\ref{CHNS1}) - (\ref{CHNS4}) and 2D MHD Eqs.~(\ref{MHD1}) - (\ref{MHD4}) with the PIXIE2D code \cite{chacon_implicit_2002,chacon_2d_2003}. The simulation box size is $L_0\times L_0=2\pi\times 2\pi$, and the resolution is $1024\times 1024$. External forcing is applied to the $A$ and $\phi$ field with the sinusoidal form $f_{A, \phi}(x,y)=f_{0A,\phi}\sin[x*\mathrm{int}(k_{fA,\phi}\cos\theta_{A,\phi})+y*\mathrm{int}(k_{fA,\phi}\sin\theta_{A,\phi})+\varphi_{A,\phi}]$, where $f_0$ is the forcing magnitude, $k_f$ is the forcing scale, and $\theta,\varphi\in[0,2\pi)$ are random angle and random phase that change at each time step, respectively. This kind of external forcing keeps the system isotropic and homogeneous.

The free parameters in the equations are $\xi$ (or $\mu_0$), $D$ (or $\eta$), $\nu$, and $\rho$. In addition, the external forcing properties $f_{0A,\phi}$, and $k_{fA,\phi}$ are also adjustable. Important dimensionless numbers here are as follows \cite{perlekar_two-dimensional_2015,pal_binary-fluid_2015}: 
\begin{itemize}
\item $L_H/L_d=Hd$, the ratio of the Hinze scale to dissipation scale.
\item $Re_\lambda=\sqrt{10}E^K/\rho\sqrt{\epsilon\nu}$, the Taylor microscale Reynolds number.
\item $Sc=\nu/D$, the Schmidt number; or $Pr=\nu/\eta$, the Prandtl number.
\item $Ch=\xi/L_0$, the Cahn number, which is the ratio of the interfacial thickness to the system size.
\item $We=\rho L_f f_{0\phi}/\sigma$ (where $L_f=2\pi/k_f$), the forcing scale Weber number, which characterizes the relative importance of the external forcing compared to the surface tension.
\item $Gr=L_0^2f_{0\phi}/\nu^2$, the Grashof number, which approximates the ratio of the external forcing to viscosity.
\end{itemize}
We keep $Sc=Pr=1$ in all our runs, and other parameters are listed in Table~\ref{parameter_table}.

\begin{figure}[htbp]
   \centering
   \includegraphics[width=0.8\columnwidth]{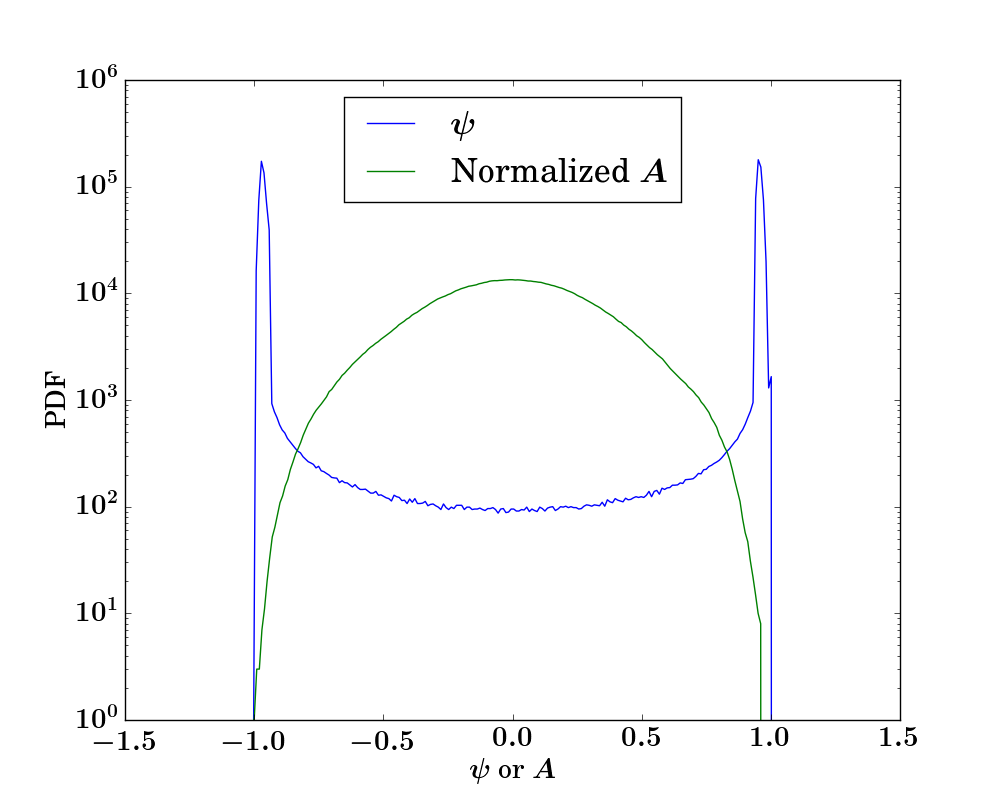} 
   \caption{The Probability Density Function (PDF) of $\psi$ (Run2) and normalized $A$ (Run6). The PDF of $\psi$ falls into the range $[-1, 1]$ spontaneously.}
   \label{pdf_of_psi}
\end{figure}

The system is periodic in both directions. The initial condition for the concentration field $\psi$ (or magnetic potential field $A$) is a random distribution of $+1$ and $-1$, while the stream function field $\phi$ is $0$ everywhere initially. Although the range of $\psi$ is $[-1, 1]$ from its physics definition $\psi = \frac{\rho_A-\rho_B}{\rho_A+\rho_B}$, we don't enforce this restriction in our simulation and let it freely evolve according to Eqs.~(\ref{CHNS1}) - (\ref{CHNS4}). This approach is valid because the Probability Density Function (PDF) of $\psi$ lies mostly in the range $[-1, 1]$ spontaneously, as shown in Fig.~\ref{pdf_of_psi}. This PDF is consistent with previous studies \cite{o_naraigh_bubbles_2007}.

2D simulations are sufficient to capture much of the important physics of the CHNS turbulence. The length scale growth, the arrest of the length scale growth, the emergence of the Hinze scale, and the inverse cascade of $H^\psi$ appear both in 3D and 2D simulations \cite{perlekar_spinodal_2014}. It is well known that 2D and 3D Navier-Stokes turbulence have totally different cascades and spectra, but 2D and 3D MHD turbulence are rather more similar. So as an analogy, 2D and 3D CHNS turbulence also should not differ much. 

\subsection{Benchmark}

\begin{figure}[htbp]
   \centering
   \includegraphics[width=0.8\columnwidth]{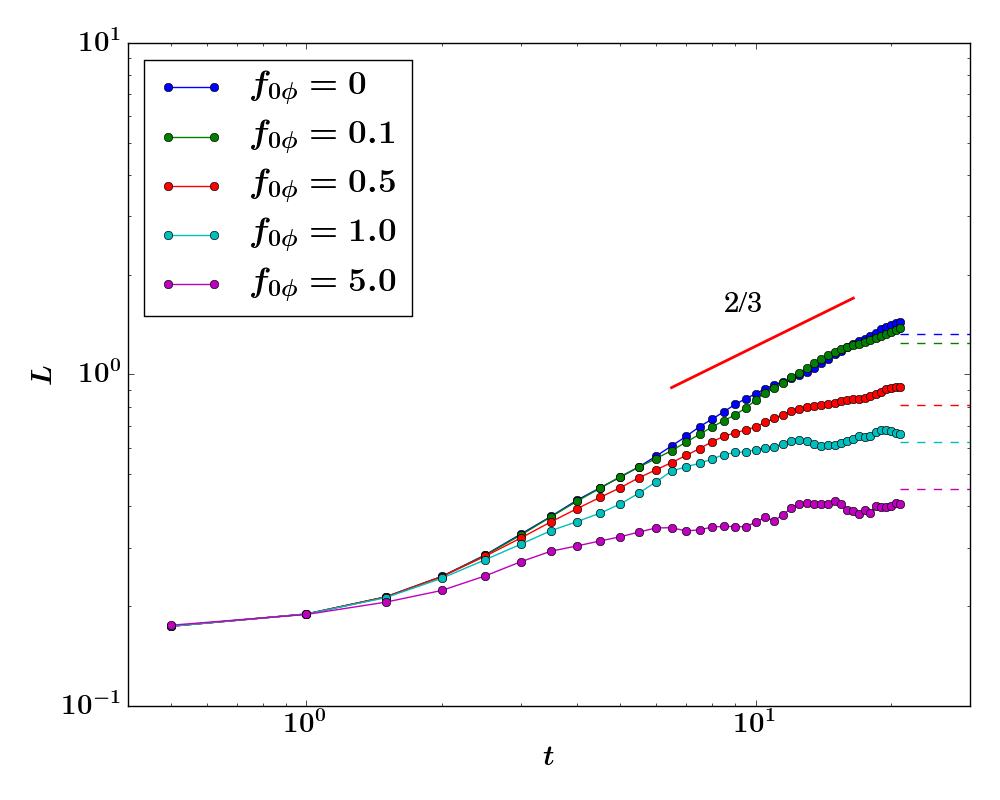} 
   \caption{Blob size growth for Run1 - Run5. Dashed lines are corresponding the Hinze scales.}
   \label{ltpng}
\end{figure}

In the simulation, we verified that, if unforced, the blob coalescence progresses, and the blob size grows until it reaches the system size. If $\phi$ field is forced at large scale, blob size growth can be arrested. See Fig.~\ref{pcolor} as an illustration. Define the blob size $L$ as the following:
\be
L(t)=2\pi \frac{\int S_k(k,t)\intd k}{\int kS_k(k,t)\intd k}\label{blob_size}
\ee
where $S_k(k,t)=\langle|\psi_\mathbf{k}(\mathbf{k},t)|^2\rangle$ is the structure function. This definition essentially picks the peak of the structure function, if it has a clear peak. 

Earlier numerical studies \cite{kendon_inertial_2001,kendon_3d_1999} observed that, if the system is unforced, the blob size $L$ grows such that $L\sim t^{2/3}$ at the late stage of the blob coalescence process. This exponent can be obtained dimensionally by balancing the advection term $\mathbf{v}\cdot\nabla\omega$ and the surface tension force term $\frac{\xi^2}{\rho}\mathbf{B}_\psi\cdot\nabla\nabla^2\psi$ in Eq.~(\ref{CHNS2}) and assuming the velocity can be estimated by $\mathbf{v}\sim\dot{L}$. The presence of external forcing can arrest the length scale growth \cite{berti_turbulence_2005}. Larger forcing leads to a larger enstrophy dissipation rate $\epsilon_\Omega$, and thus a smaller Hinze scale. Fig.~\ref{ltpng} supports this finding. The peak of the $H^\psi_k$ spectrum moving towards larger scale in Fig.~\ref{Spectra_with_time} is consistent with the blob size $L$ growth shown in Fig.~\ref{ltpng}. 

\subsection{The $H^\psi_k$ Flux}

\begin{figure*}[htbp]
   \centering
   \includegraphics[width=0.8\textwidth]{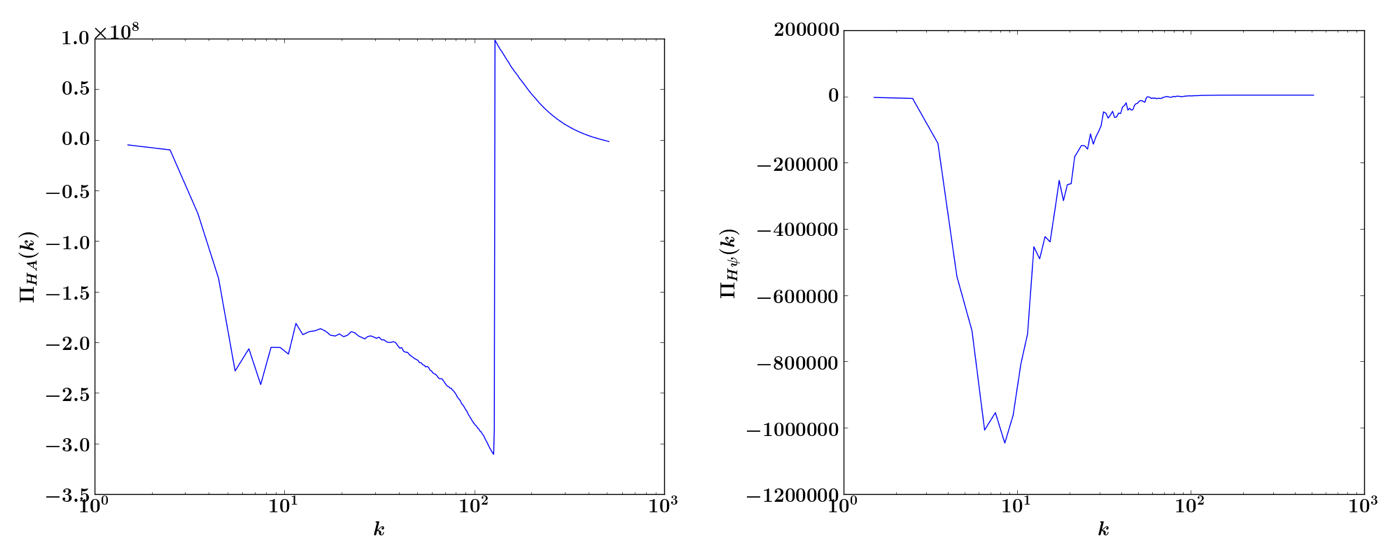} 
   \caption{The $H^A_k$ flux (left) for MHD (Run6), and the $H^\psi_k$ flux (right) for CHNS (Run2). }
   \label{flux}
\end{figure*}

The directions of cascades are suggested by the sign of the corresponding spectral fluxes. We define the $H^\psi_k$ flux and the $H^A_k$ flux as follows:

\be
\Pi_{HA}(k)=\sum_{k<k'}T_{HA}(k')\text{, where }T_{HA}(k)=\langle A_k^*(\mathbf{v}\cdot\nabla A)_k\rangle
\ee

\be
\Pi_{H\psi}(k)=\sum_{k<k'}T_{H\psi}(k')\text{, where }T_{H\psi}(k)=\langle \psi_k^*(\mathbf{v}\cdot\nabla \psi)_k\rangle
\ee

If a flux is negative, then the corresponding transfer is inverse, suggestive of an inverse cascade. See Fig.~\ref{flux} for our simulation results. For the MHD case (left), an external forcing on the magnetic potential $A$ is applied on $k=128$. The small scale $A$ forcing drives an inverse transfer of $H^A$. For the CHNS case (right), no forcing on $\psi$ is necessary for the appearance of an inverse transfer of $H^\psi$. The negative diffusion term in the CHNS equations leads to small scale instability. Thus it plays a similar role to forcing of $\psi$.

\subsection{The $H^\psi_k$ Spectrum Power Law}

\begin{figure*}[htbp]
   \centering
   \includegraphics[width=0.8\textwidth]{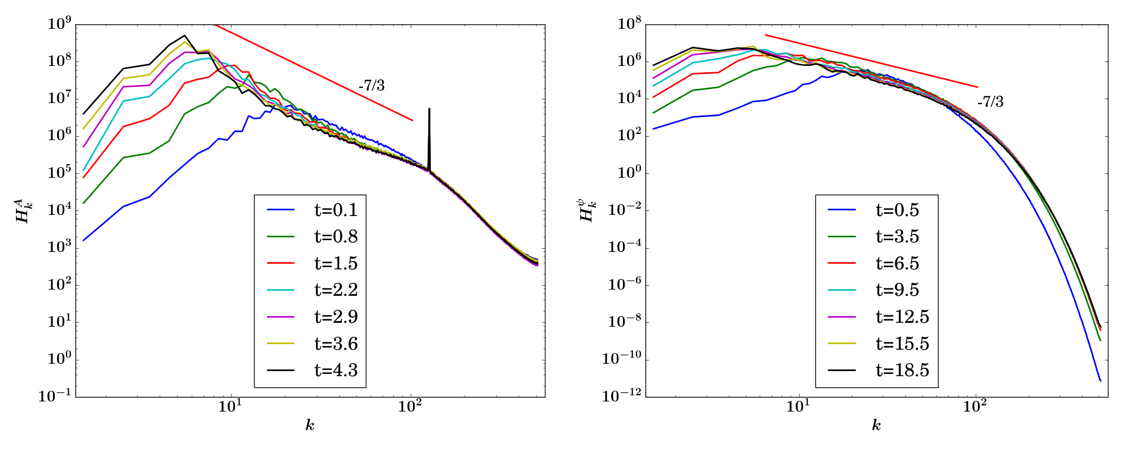} 
   \caption{The $H^A_k$ spectrum in 2D MHD for Run6 at various times (left), and the $H^\psi_k$ spectrum in 2D CHNS for Run2 (right).}
   \label{Spectra_with_time}
\end{figure*}

It is known that the dynamics of 2D MHD turbulence is dominated by the inverse cascade of $H^A$, if $H^A$ is injected at small scales. The corresponding power law of the $H^A_k$ spectrum is $-7/3$:
\be
H^A_k\sim\epsilon_{HA}^{2/3}k^{-7/3}\label{HA}
\ee
Here $\epsilon_{HA}$ is the $H^A$ dissipation rate, and see Fig.~\ref{Spectra_with_time} (left) for the simulation result. Note that in order to obtain a 2D MHD setup similar to the 2D CHNS system, small scale external forcing of the $A$ field and large scale external forcing of the $\phi$ field are imposed. 

The scaling argument for the power of $-7/3$ for 2D MHD is as follows. Assuming there is a constant mean square magnetic potential dissipation rate $\epsilon_{HA}$, according to the Alfvenic equipartition ($\rho\langle \mathbf{v}^2 \rangle \sim \frac{1}{\mu_0}\langle \mathbf{B}^2 \rangle$), the time scale for the decay of $H^A$ ($\epsilon_{HA}\sim H^A/\tau$) can be estimated by $\tau\sim (v^{rms}k)^{-1} \sim (B^{rms}k)^{-1}$. Define the spectrum to be $H^A=\sum_k H^A_k\sim kH^A_k$, so $B^{rms}\sim kA\sim k(H^A)^{1/2}\sim(H^A_k)^{1/2}k^{3/2}$. Therefore, $\epsilon_{HA}\sim H^A/\tau\sim (H^A_k)^{2/3}k^{7/2}$, leading to Eq.~(\ref{HA}).

\begin{figure}[htbp]
   \centering
   \includegraphics[width=0.8\columnwidth]{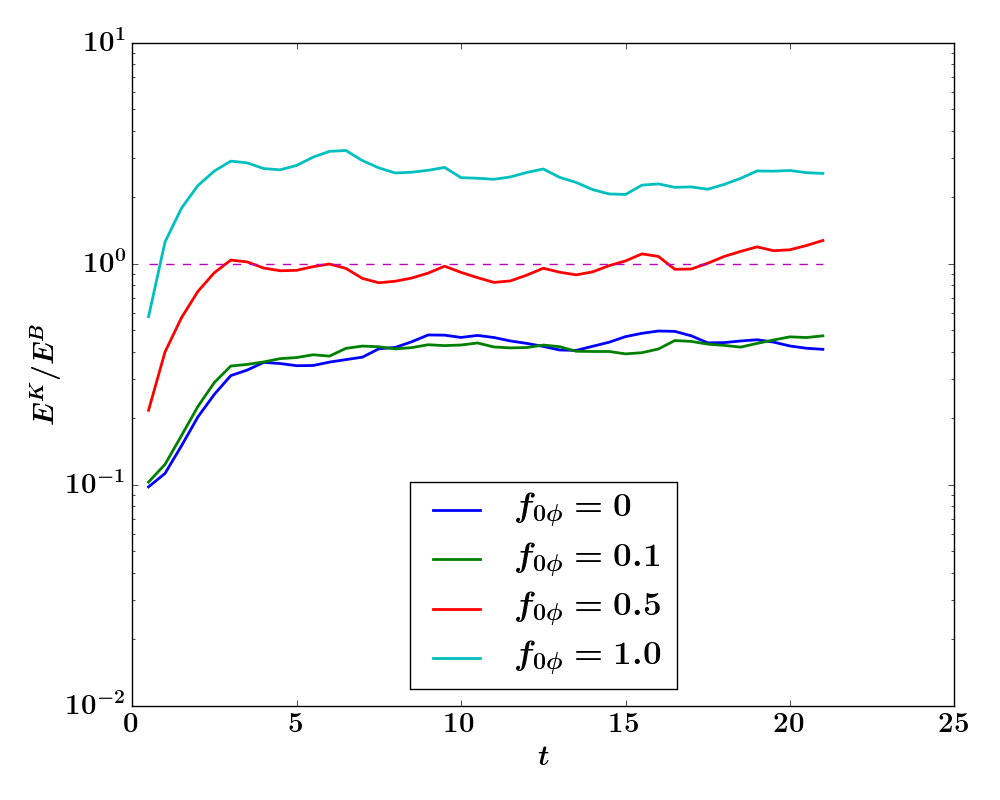} 
   \caption{The ratio of $E^K$ to $E^B$ for Run1 - Run4 supports the assumption of elastic equipartition ($\rho\langle \mathbf{v}^2 \rangle \sim \xi^2\langle \mathbf{B}_\psi^2 \rangle$). If the forcing intensity is too strong, then the elastic forcing term in the $\omega$ equation becomes negligible, and the system does not significantly differ from the 2D NS equation. In our study, though we tried a broad range of forcing intensity, larger forcing (than Run5) may break the equipartition of the kinetic and magnetic energy.}
   \label{equipartition}
\end{figure}

The same argument can be applied to 2D CHNS turbulence to get a (similar) $H_\psi$ spectrum. Assuming that elastic equipartition applies to the 2D CHNS system ($\rho\langle \mathbf{v}^2 \rangle \sim \xi^2\langle \mathbf{B}_\psi^2 \rangle$) (see Fig.~\ref{equipartition}), the time scale for the decay of $H^\psi$ is $\tau\sim (v^{rms}k)^{-1} \sim (B_\psi^{rms} k)^{-1}$. Then by repeating the above argument for MHD, it is easy to obtain the $H^\psi_k$ spectrum:
\be
H^\psi_k\sim\epsilon_{H\psi}^{2/3}k^{-7/3}\label{psi73}
\ee

The simulation result for the $H^\psi_k$ spectrum in 2D CHNS turbulence in Fig.~\ref{Spectra_with_time} (right) verifies the similarity to the $H^A_k$ spectrum in 2D MHD turbulence. The peak of the $H^\psi_k$ spectrum, which gives the approximate blob size according to Eq.~(\ref{blob_size}), moves towards larger scale, as shown in Fig.~\ref{Spectra_with_time}. The blob coarsening process is consistent with the inverse cascade of $H^\psi$. Moreover, the $H^\psi_k$ spectrum with power law $-7/3$ is indeed a good fit, as predicted by the inverse cascade of $H^\psi$ argument. Again, we assumed (marginally satisfied) elastic equipartition in order to obtain the $-7/3$ power law. The result fits the simulation very well. These findings suggest that the dynamics of the fluctuating concentration field is governed by the inverse cascade of $H^\psi$.

\begin{figure}[htbp]
   \centering
   \includegraphics[width=0.8\columnwidth]{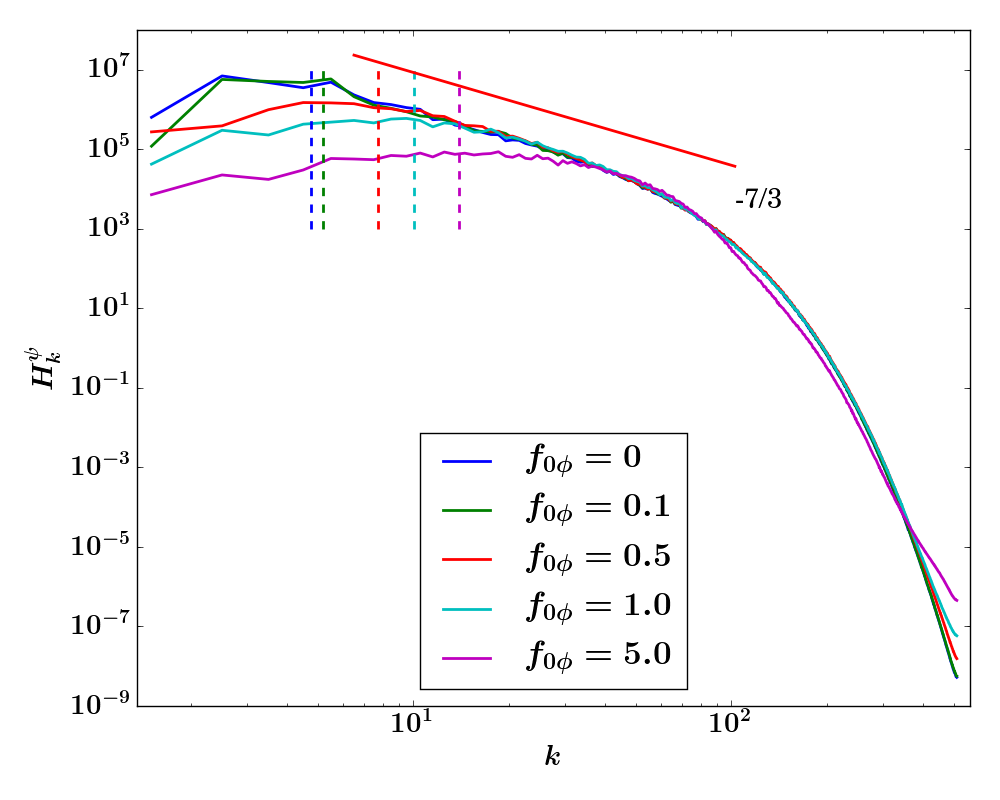} 
   \caption{$H^\psi$ spectra for Run1 - Run4, with different magnitudes of external forcing $f_{0\phi}$ thus different Hinze scales. The Hinze scale for each run is marked by a dashed line with the same color.}
   \label{Spectra_with_f}
\end{figure}

The $-7/3$ power is robust. It does not change with the magnitude of external forcing, as long as the separation between the Hinze scale and the dissipation scale is maintained, so the elastic range is long enough ($Hd\gg 1$). Fig.~\ref{Spectra_with_f} gives the $H^\psi_k$ spectra for different external forcing strengths. It shows that the power $-7/3$ remains unchanged. Note that larger external forcing leads to a smaller Hinze scale according to Eq.~(\ref{Hinze_dir}), so the elastic range is shorter. If the Hinze scale is close to or even smaller than the dissipation scale, there will be no clear elastic range, and thus no power law spectrum for $H^\psi_k$. Thus, a sufficient separation between the Hinze scale and the dissipation scale ($Hd\gg 1$) is critical to uncovering elastodynamic phenomena.

\subsection{The Energy Spectrum Power Law}

\begin{figure*}[htbp]
   \centering
   \includegraphics[width=0.8\textwidth]{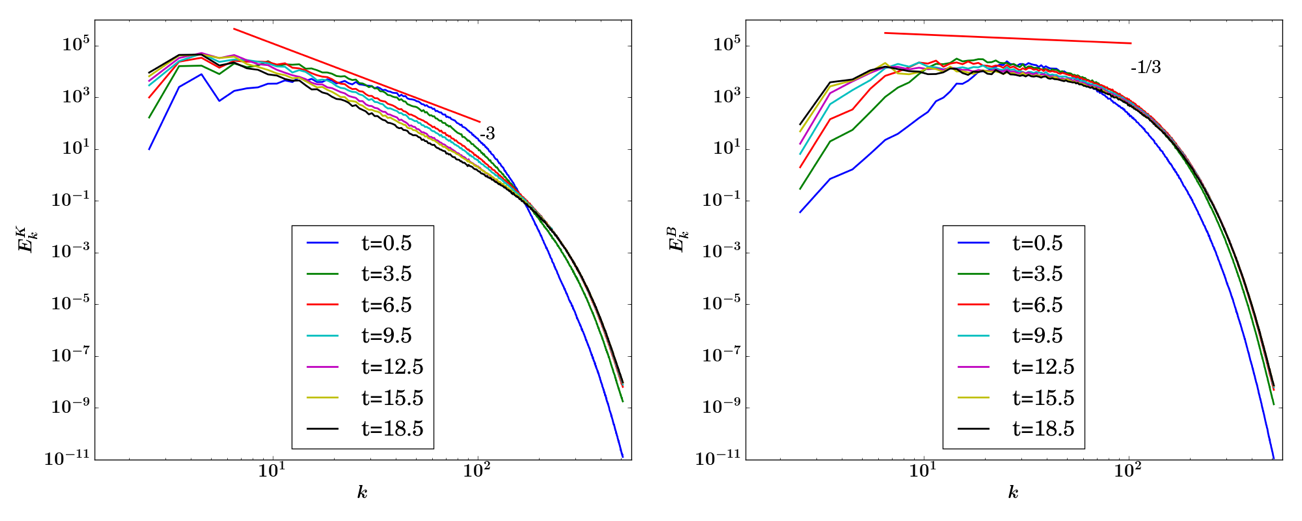} 
   \caption{Kinetic energy spectrum (left) and magnetic energy spectrum (right) for Run2. The kinetic energy spectrum indicates a direct enstrophy cascade of 2D NS turbulence.}
   \label{EkEb}
\end{figure*}

When the $\phi$ field is forced at large scale, the kinetic energy spectrum is $E^K_k\sim k^{-3}$, as shown in Fig.~\ref{EkEb}. This spectrum is the same as that for the direct enstrophy cascade in 2D NS Turbulence. This result is initially surprising, because enstrophy is \textit{not} a conserved quantity in the 2D CHNS system. The kinetic energy spectrum for 2D CHNS turbulence is different from that for 2D MHD turbulence. It is well known that in the direct energy cascade regime of 2D MHD, the energy spectrum is $E^K_k\sim k^{-3/2}$, which is called the Iroshnikov-Kraichnan (IK) spectrum \cite{kraichnan_inertial-range_1965,iroshnikov_turbulence_1964}. The IK spectrum is the consequence of the interaction between Alfven waves propagating in opposite directions. The result that the kinetic energy spectrum for the 2D CHNS system is significantly different from the IK spectrum for MHD suggests that the back reaction of surface tension on the fluid motion is limited. 

\begin{figure}[htbp]
   \centering
   \includegraphics[width=0.8\columnwidth]{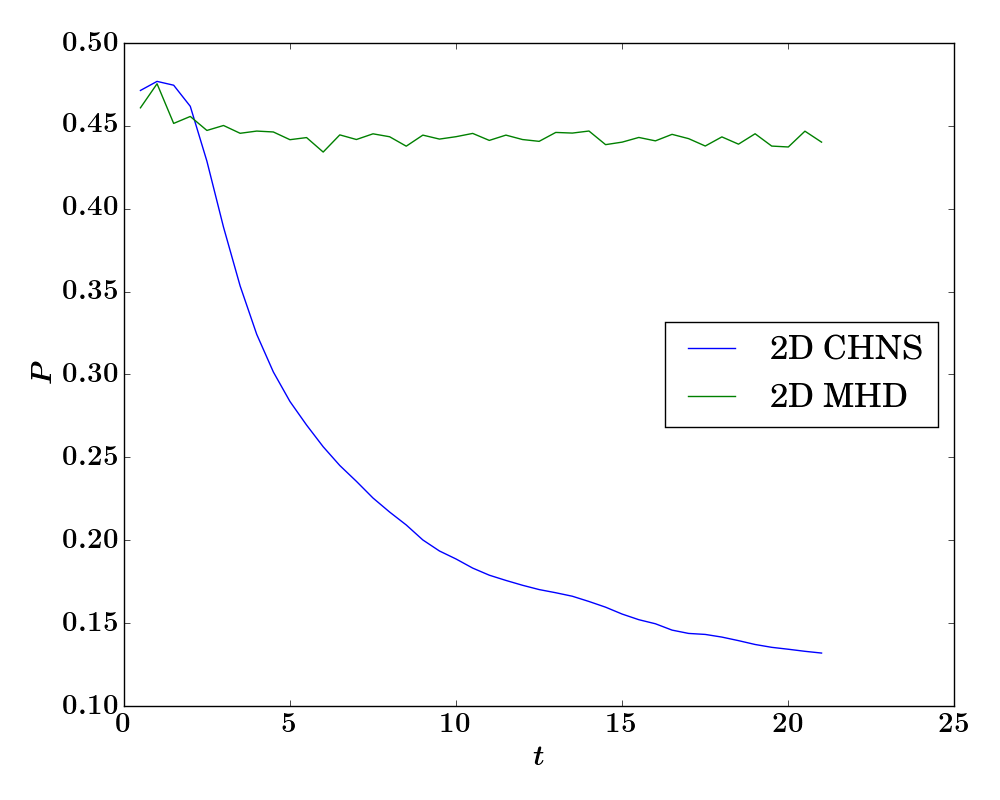} 
   \caption{The time evolution for the interface packing fraction $P$, the ratio of mesh grid number where $|\mathbf{B}_\psi|>B_\psi^{rms}$ (or $|\mathbf{B}|>B^{rms}$) over total mesh grid number. }
   \label{Rtime}
\end{figure}

\begin{figure}[htbp]
   \centering
   \includegraphics[width=\columnwidth]{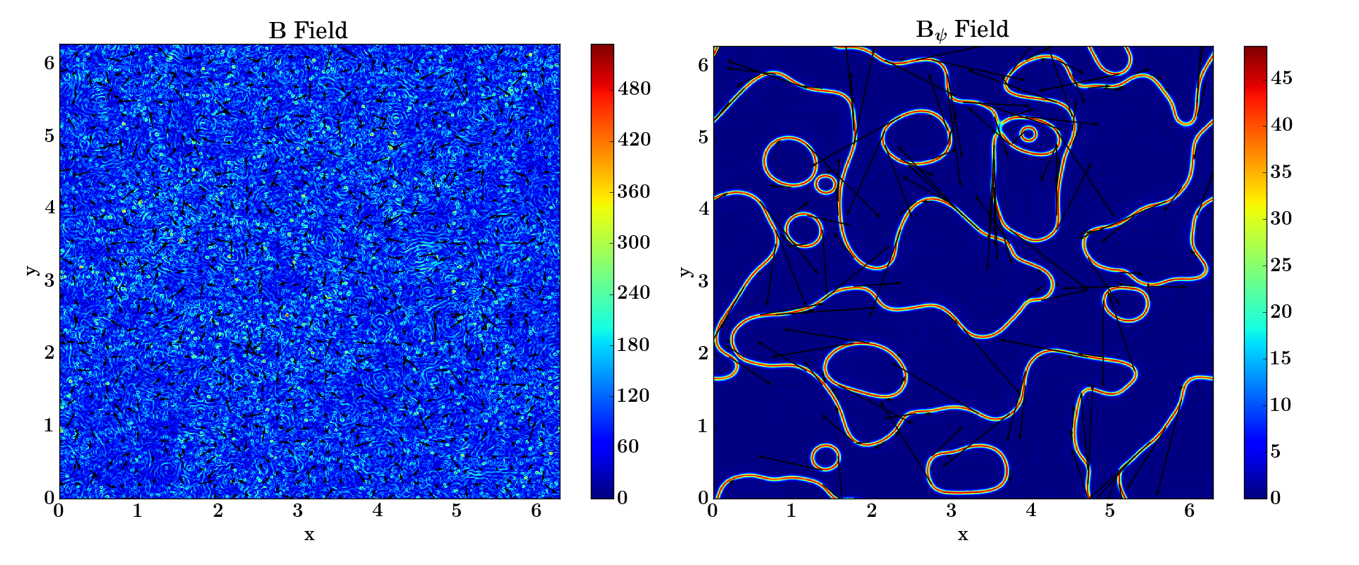} 
   \caption{$\mathbf{B}$ field for Run6 (left) and $\mathbf{B}_\psi$ field for Run2 (right). From the color map we can see that the structures look quite different.}
   \label{bfield}
\end{figure}

This initially surprising result is plausible because in the 2D CHNS system, $\mathbf{B}_\psi$ vanishes in most of the space. $\mathbf{B}_\psi$ is large only in the interfacial regions, and the interfacial regions fill only a small portion of the system, as shown in Fig.~\ref{bfield}. On the other hand, the magnetic fields in MHD are not localized to specific regions, so Alfven waves can propogate everywhere. Define the interface packing fraction $P$ to be the ratio of mesh grid number where $|\mathbf{B}_\psi|>B_\psi^{rms}$ (or $|\mathbf{B}|>B^{rms}$) to the total mesh grid number. This definition of interface packing fraction is a rather simple choice of a figure of merit, but one for which we can easily grasp the underlying physics. In the 2D CHNS system, $P=13.9\%$ for Run2; while for 2D MHD, $P=44.0\%$ for Run6. This notable difference shows that only a small portion of the 2D CHNS system is strongly affected by the $\mathbf{B}_\psi$ field, as compared to MHD. The time evolution for the interface packing fraction $P$ is shown in Fig.~\ref{Rtime}. In the 2D CHNS system, as time progresses, the blob coalescence process drives the interfacial region to a smaller and smaller interface packing fraction, and thus suppresses the elastic effects on fluid motion. If there is a larger number of blobs, there will be a larger interfacial region, and thus the velocity field will be more heavily influenced by the $\mathbf{B}_\psi$ field. In that case, the kinetic energy spectrum will be more MHD-like.

\section{Conclusion and Discussion}\label{sec6}

\begin{table*}
\caption{Contrast of 2D MHD and the 2D CHNS system.}
\begin{center}
\begin{tabular}{ccc}
\hline
\hline
& 2D MHD & 2D CHNS\\
\hline
Diffusion & A simple positive diffusion term & A negative, a self nonlinear, and a hyper-diffusion term\\
Range of potential & No restriction for range of $A$ & $\psi\in[-1,1]$ \\
Interface Packing Fraction & Not far from $50\%$ & Small\\
Back reaction & $\mathbf{j}\times\mathbf{B}$ force can be significant & Back reaction is apparently limited\\
Kinetic energy spectrum & $E^K_k\sim k^{-3/2}$ & $E^K_k\sim k^{-3}$\\
Suggestive cascade by $E^K_k$ & Suggestive of direct energy cascade & Suggestive of direct enstrophy cascade\\
\hline
\hline
\end{tabular}
\end{center}
\label{contrast}
\end{table*}

2D CHNS turbulence is an analogue to 2D MHD turbulence. The two systems have some common features and also some important differences. See Table~\ref{comparison} for comparison and Table~\ref{contrast} for contrasts. The theories of 2D MHD turbulence give us inspiration and guidance for the study of 2D CHNS turbulence.

From the basic equations, it is easy to notice similarities between 2D CHNS and 2D MHD. Most clear is that the surface tension force is a direct analogue of the $\mathbf{j}\times\mathbf{B}$ force. The ideal quadratic conserved quantities of these two systems have the same form, and this leads to the same cascade directions. The linear elastic wave from the 2D CHNS system has a similar dispersion relation to the Alfven wave from 2D MHD. The linear elastic wave plays an important role in the dynamics through the elasticization process, which is analogous to the Alfvenization process.

The scales between the Hinze scale and dissipation scale in the 2D CHNS system form the elastic range. Separation of the Hinze scale and the dissipation scale ($Hd\gg 1$) is critical to allow an elastic range. In the elastic range, the surface tension interaction induces an elastic effect critical to the nonlinear dynamics, so the system is more MHD-like.

By direct numerical simulation, we find that in the elastic range, the mean square concentration spectrum is $H^\psi_k\sim k^{-7/3}$. This power law scaling can be recovered theoretically by assuming elastic equipartition (which is at best marginally satisfied). The $-7/3$ power law is the same as the $H^A_k$ spectrum in the inverse cascade regime of 2D MHD. The $-7/3$ power law is robust and independent of the forcing strength. This result suggests that the dynamics of the fluctuating concentration field is governed by the inverse cascade of $H^\psi_k$. The inverse cascade of $H^\psi$ is consistent with the blob coalescence process.

The kinetic energy spectrum for the 2D CHNS system is $E^K_k\sim k^{-3}$ when forced at large scale. This spectrum is different from the IK spectrum in MHD, and is the same as the kinetic energy spectrum in the 2D NS turbulence direct enstrophy cascade regime. This result suggests that the back reaction of surface tension on the fluid motion is limited. This is plausible because the back reaction is only significant in the interfacial regions, which fill only a small part of the system. This is an important difference between 2D CHNS turbulence and 2D MHD turbulence. In order to make the kinetic energy spectrum more MHD-like, we need to increase the interface packing fraction. We will obtain larger interfacial regions if we have a large number of small blobs instead of a small number of large blobs. Thus the apparent next step is to increase the forcing strength or change the form of forcing in order to increase the interface packing fraction. However, a larger forcing strength leads to a smaller Hinze scale, and thus a shorter elastic range. If we want to keep a broad enough elastic range, we have to decrease the dissipation scale at the same time, i.e. decrease $\nu$. This requires higher resolution and more computing resources, and so we will perform runs with higher resolution in future works. The definition of interface packing fraction we use in this paper is rather crude, and more study about how to characterize the interface, what physics controls the interface packing fraction, and how to increase the interface packing fraction would be interesting. The statistics of $|\mathbf{B}_\psi|$ and how it is related to the interface packing fraction is also a relevant interesting problem to study.

The theories of 2D MHD turbulence can also inspire the study of turbulent transport and memory effects in 2D CHNS turbulence. Even a weak mean magnetic field can result in a large mean square fluctuation. Such small scale magnetic fields will result in enhanced memory, so turbulent transport in MHD with even a weak large scale magnetic field is suppressed \cite{cattaneo_suppression_1991,tobias_-plane_2007,vainshtein_nonlinear_1992,diamond_self-consistent_2005}. This effect may also appear in 2D CHNS turbulence. It is also interesting to investigate the possible change of momentum transport in the elastic range of CHNS, due to elastic wave effects. 2D CHNS turbulence also has similarities to elastic turbulence in polymer solutions \cite{tabor_cascade_1986,de_gennes_towards_1986}. The comparison and contrast among MHD, CHNS and polymer hydrodynamic turbulence will be discussed in future works.

\begin{acknowledgments}
We thank David Hughes and Steve Tobias for useful conversations. P. H. Diamond thanks Annick Pouquet for a fascinating discussion of competing cascades in MHD turbulence. We acknowledge the hospitality of Peking University where part of this research was performed. Xiang Fan thanks Los Alamos National Laboratory for its hospitality and help with computing resources. We thank the participants at the 2015 Festival de Th\'eorie for discussions and comments. This research was supported by the U.S. Department of Energy, Office of Science, Office of Fusion Energy Sciences, under Award Number DE-FG02-04ER54738 and CMTFO.
\end{acknowledgments}

\begin{table*}
\caption{Simulation parameters. Note that for 2D MHD runs, $\xi$ means $\mu_0^{-1/2}$, and $D$ means $\eta$.}
\scriptsize
\begin{center}
\begin{tabular}{p{0.05\linewidth}p{0.05\linewidth}p{0.05\linewidth}p{0.05\linewidth}p{0.05\linewidth}p{0.05\linewidth}p{0.05\linewidth}p{0.05\linewidth}p{0.05\linewidth}p{0.05\linewidth}p{0.05\linewidth}p{0.05\linewidth}p{0.05\linewidth}p{0.06\linewidth}p{0.08\linewidth}}
\hline
\hline
Run & System & $\xi$ & $D$ & $\nu$ & $\rho$ & $f_{0\phi}$ & $k_{f\phi}$ & $f_{0A}$ & $k_{fA}$ & $Re_\lambda$ & $Hd$ & $We$ & $Gr$ & $\xi^2/\rho$\\
\hline
Run1 & CHNS & $0.015$ & $10^{-3}$ & $10^{-3}$ & $1.0$ & $0$ & $-$ & $-$ & $-$ & $5.5$ & $39$ & $0$ & $0$ & $2.25*10^{-4}$\\ 
Run2 & CHNS & $0.015$ & $10^{-3}$ & $10^{-3}$ & $1.0$ & $0.1$ & $4$ & $-$ & $-$ & $6.1$ & $39$ & $11$ & $3.9*10^6$ & $2.25*10^{-4}$\\ 
Run3 & CHNS & $0.015$ & $10^{-3}$ & $10^{-3}$ & $1.0$ & $0.5$ & $4$ & $-$ & $-$ & $25$ & $35$ & $56$ & $2.0*10^7$ & $2.25*10^{-4}$\\ 
Run4 & CHNS & $0.015$ & $10^{-3}$ & $10^{-3}$ & $1.0$ & $1.0$ & $4$ & $-$ & $-$ & $59$ & $33$ & $110$ & $3.9*10^7$ & $2.25*10^{-4}$\\ 
Run5 & CHNS & $0.015$ & $10^{-3}$ & $10^{-3}$ & $1.0$ & $5.0$ & $4$ & $-$ & $-$ & $719$ & $30$& $550$ & $2.0*10^8$ & $2.25*10^{-4}$\\ 
Run6 & MHD & $0.015$ & $10^{-3}$ & $10^{-3}$ & $1.0$ & $1.0$ & $4$ & $10^3$ & $128$ & $18$ & - & - & - & $2.25*10^{-4}$\\ 
\hline
\hline
\end{tabular}
\end{center}
\label{parameter_table}
\end{table*}%

\bibliography{spectrum_paper}

\end{document}